\documentclass[journal,a4paper]{IEEEtran}
\usepackage[utf8]{inputenc}
\usepackage{amsmath}
\usepackage{mathtools}
\usepackage{graphicx}
\usepackage{float}
\usepackage{pgfplots}
\usepackage{tikz}
\usetikzlibrary{plotmarks}
\usetikzlibrary{external}
\usetikzlibrary{arrows}
\usetikzlibrary{shapes,decorations}
\usetikzlibrary{colorbrewer}
\usepackage{caption}
\usepackage{cite}
\pgfplotsset{compat=1.14}
\usepackage{xcolor}
\usetikzlibrary{shapes}
\usetikzlibrary{plotmarks}
\usetikzlibrary{external}
\usetikzlibrary{arrows}
\usetikzlibrary{shapes,decorations}
\usetikzlibrary{colorbrewer}
\usepackage{caption}
\pgfplotsset{compat=1.14}
\usepackage[utf8]{inputenc}

\DeclarePairedDelimiter\ceil{\lceil}{\rceil}
\newcommand\ninf[1]{\left\lfloor#1\right\rceil}

\usepackage{amssymb}

\begin{document}

\title{The Impact of Transceiver Noise on Digital Nonlinearity Compensation}

\author{Daniel~Semrau,~\IEEEmembership{Student Member,~IEEE,}   Domani\c{c}~Lavery,~\IEEEmembership{Member,~IEEE,} Lidia Galdino,~\IEEEmembership{Member,~IEEE,}
        Robert~I.~Killey,~\IEEEmembership{Senior Member,~IEEE,}
        and~Polina~Bayvel,~\IEEEmembership{Fellow,~IEEE,~Fellow,~OSA}
\thanks{This work was supported by a UK EPSRC programme grant UNLOC (EP/J017582/1) and a Doctoral Training Partnership (DTP) studentship for Daniel Semrau. D.~Lavery is supported by the Royal Academy of Engineering under the Research Fellowships scheme.}
\thanks{D. Semrau, D. Lavery, Lidia Galdino, R. I. Killey, and P.
Bayvel are with the Optical Networks Group, University College London, London
WC1E 7JE, U.K. (e-mail: \{uceedfs; d.lavery; l.galdino; r.killey; p.bayvel\}@ucl.ac.uk.)}
}

\maketitle

\markboth{\today}%
{}

\begin{abstract}
The efficiency of digital nonlinearity compensation (NLC) is analyzed in the presence of noise arising from amplified spontaneous emission noise (ASE) as well as from a non-ideal transceiver subsystem. Its impact on signal-to-noise ratio (SNR) and reach increase is studied with particular emphasis on split NLC, where the digital back-propagation algorithm is divided between transmitter and receiver. An analytical model is presented to compute the SNR's for non-ideal transmission systems with arbitrary split NLC configurations. When signal-signal nonlinearities are compensated, the performance limitation arises from residual signal-noise interactions. These interactions consist of nonlinear beating between the signal and co-propagating ASE and transceiver noise. While transceiver noise-signal beating is usually dominant for short transmission distances, ASE noise-signal beating is dominant for larger transmission distances. It is shown that both regimes behave differently with respect to the optimal NLC split ratio and their respective reach gains. Additionally, simple formulas for the prediction of the optimal NLC split ratio and the reach increase in those two regimes are reported. 
It is found that split NLC offers negligible gain with respect to conventional digital back-propagation (DBP) for distances less than 1000~km using standard single-mode fibers and a transceiver (back-to-back) SNR of 26~dB, when transmitter and receiver inject the same amount of noise. However, when transmitter and receiver inject an unequal amount of noise, reach gains of 56\% on top of DBP are achievable by properly tailoring the split NLC algorithm. The theoretical findings are confirmed by numerical simulations.
\end{abstract}


\IEEEpeerreviewmaketitle

\section{Introduction}

\IEEEPARstart{D}{igital} nonlinearity compensation (NLC) offers a great potential in overcoming the limit in optical communication systems imposed by fiber nonlinearity \cite{Ip_2010_nlu,Liga_2014_otp,Semrau_2016_air}. Most digital nonlinearity compensation techniques extend the physical link with a virtual link in the digital signal processing (DSP) stage using an inverted propagation equation. To date, three different implementations have been proposed in the literature, depending on whether this virtual link is placed at the transmitter, receiver or evenly split between them.
\par
Receiver-side NLC, also called digital back-propagation (DBP), has been proposed in numerous research papers to reduce the impact of fiber nonlinearities and achieve improved transmission performance
\cite{Ip_2008_cod,Millar_2010_mof,Irukulapati_2014_sdb,Tanimura_2012_aro,Fontaine_2013_fnc,Dar_2016_onl}. Reach increases of around 100\% (from $640$~km to $1280$~km) and 150\% (from $1000$~km to $2500$~km) have been experimentally demonstrated, when NLC is applied jointly to all received channels\cite{Maher_2015_reo,Galdino_2017_otl}. For shorter distances, even a threefold increase in transmission distance was experimentally achieved (233\% from $300$~km to $1000$~km) \cite{Galdino_2017_otl}. Overcoming the relatively small bandwidths of digital-to-analog converters and the use of mutually coherent sources enabled the application of transmitter-side NLC, sometimes referred to as transmitter-side DBP or digital precompensation (DPC) \cite{Temprana_2016_tdb}. Reach gains of 100\% (from $1530$~km to $3060$~km) and 200\% (from $425$~km to $1275$~km) for shorter distances have been shown experimentally \cite{Temprana_2015_ttr,Temprana_2016_doc}. 
\par 
The performance difference between transmitter-side and receiver-side NLC lies only in the periodic arrangement of the optical amplifiers along the link \cite{Lavery_2017_otb}. This is due to over-/under-compensated ASE noise-signal interactions (hereafter ``ASE noise beating'') that strongly depend on the specific location where each ASE noise contribution is introduced. For conventional links, where an optical amplifier is located after each span, DPC improves the transmission performance by up to one additional span. The gain in signal-to-noise ratio (SNR) decreases with distance and is approximately $0.2$~dB after $20$ spans and less than $0.1$~dB after more than $45$ spans \cite{Lavery_2016_tbo}.
\par
Apart from transmitter and receiver-side NLC, an implementation has been proposed where the virtual link is equally divided between transmitter and receiver, which is referred to as split NLC or split DBP \cite{Lavery_2016_tbo,Lavery_2017_otb,Ellis_2015_clo}. This approach minimizes the residual ASE noise beating and yields at least $1.5$~dB improvement in SNR compared to conventional DBP, assuming full-field compensation (NLC applied jointly to all channels) and the absence of transceiver noise. However, to date there is no experimental demonstration of this potentially advantageous scheme.
\par
All theoretical considerations above only assume ASE noise injected by optical amplifiers. Therefore, they might not generally apply for real transmission systems that further exhibit noise originating from non-ideal transceivers. Transceiver noise (TRX noise) is related to the back-to-back performance; that is, the maximal achievable SNR in a transmission system. This phenomenological quantity combines all noise contributions from transmitter and receiver such as quantization noise of analogue-to-digital (ADC) or digital-to-analogue (DAC) converters and noise from linear electrical amplifiers and optical components. We recently showed that resulting TRX noise-signal interactions (hereafter ``TRX noise beating'') significantly reduce the gains of (receiver-side) digital back-propagation \cite{Galdino_2017_otl}. 
Due to the adverse impact of transceiver noise-signal beating, the performance analysis of transmitter-side, receiver-side and split NLC must be substantially revised for practical transmission systems with realistic transceiver sub-systems. 
\par 
In this paper, digital nonlinearity compensation in the presence of transceiver noise is studied with particular emphasis on split NLC. We refer to split NLC as an \textit{arbitrary} split of the virtual link between transmitter and receiver, including DPC and DBP as special cases. The contributions of the paper are twofold. First, an analytical model is presented that predicts the received SNR for any arbitrary NLC split ratio (in Sec. \ref{sec:theory}). Second, using this model, two regimes are defined depending on the negligibility of either TRX or ASE noise beating. The two regimes are studied separately as they both exhibit a different behavior with respect to the optimal NLC split ratio and the achievable reach gain. Simple expressions for their computations are reported and the implications of split NLC for realistic systems are deduced (in Sec. \ref{sec:TRX_beat_regime} and \ref{ASE noise beating regime}). Finally, the theoretical findings are confirmed by numerical simulations for two cases (in Sec. \ref{sim}): One where the transmitter and receiver introduce an equal amount of noise and another where an unequal amount of noise is introduced.
\section{Analytical Model}
\label{sec:theory}
\begin{figure}[h]
\includegraphics[]{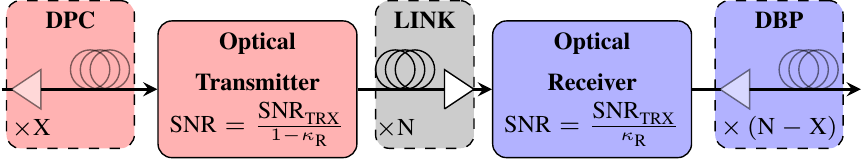}
\caption{The transmission model used for investigating the performance of fiber
nonlinearity compensation, where the digital nonlinearity compensation is
arbitrary divided between transmitter and receiver.}
\label{fig:scheme}
\end{figure}

In this section the impact of transceiver noise on split nonlinearity compensation is studied analytically. Split NLC means that nonlinearity compensation is performed over $X$ spans at the transmitter and over the remaining $N-X$ spans at the receiver, where $N$ is the total number of spans in the physical link. The transceiver noise resulting from a limited transceiver SNR (i.e., the back-to-back SNR) is divided between transmitter and receiver according to a ratio $\kappa_{\text{R}}$, where $\kappa_{\text{R}}=0$ means that all the TRX noise is injected at the transmitter and $\kappa_{\text{R}}=1$ means that all the TRX noise is injected at the receiver. In other words, the  transmitter imposes a maximum achievable SNR of $\text{SNR}_{\text{TX}}=\frac{\text{\scriptsize SNR}_{\text{TRX}}}{1-\kappa_{\text{R}}}$ and the receiver imposes a maximum achievable SNR of $\text{SNR}_{\text{RX}}=\frac{\text{\scriptsize SNR}_{\text{TRX}}}{\kappa_{\text{R}}}$, where $\text{SNR}_{\text{TRX}}$ is the transceiver SNR. The transmission set-up is schematically illustrated in Fig. \ref{fig:scheme}. 
\par 
The SNR after full-field nonlinearity compensation is given by \cite[Eq.~(6)]{Galdino_2017_otl}
\begin{align}
\text{SNR}=\frac{P}{\kappa{}P+NP_{\textnormal{ASE}}+3\eta\left(\kappa\xi_{\textnormal{TRX}}P+\xi_{\textnormal{ASE}}P_{\textnormal{ASE}}\right)P^2},
\label{eq:SNR_general}
\end{align}
where $P$ is the launch power per channel, $\kappa=\frac{1}{\text{SNR}_{\text{TRX}}}$, $P_{\text{ASE}}$ is the ASE noise per amplifier, $\eta$ is the nonlinear interference coefficient for one span, $\xi_{\textnormal{TRX}}$ is the TRX noise beating accumulation factor and $\xi_{\textnormal{ASE}}$ is the ASE noise beating accumulation factor. The latter two quantities represent uncompensated nonlinear mixing products between signal and noise and are the only quantities in Eq. \eqref{eq:SNR_general} that depend on $X$. Therefore, their minimization plays a fundamental role for split nonlinearity compensation. Both accumulation factors are given by
\begin{align}
\xi_{\textnormal{TRX}}=\underbrace{\left(1-\kappa_{\text{R}}\right)X^{1+\epsilon}}_{\textnormal{TX beat.}}+\underbrace{\kappa_{\text{R}}\left(N-X\right)^{1+\epsilon}}_{\textnormal{RX beat.}},
\label{eq:xitrx}
\end{align}
\begin{align}
\xi_{\textnormal{ASE}}=\sum_{i=1}^{X-1}i^{1+\epsilon}+\sum_{i=1}^{N-X}i^{1+\epsilon},
\label{eq:xiase}
\end{align}
where $\epsilon$ is the coherence factor \cite{Poggiolini_2012_tgm,Semrau_2017_ace}. The coherence factor is a measure for coherent accumulation of nonlinearity along the spans of a link. The first term in \eqref{eq:xitrx} represents the residual uncompensated beating between signal and transmitter noise and the second term in \eqref{eq:xitrx} represents the residual beating between signal and receiver noise. Eq. \eqref{eq:xiase} represents the residual beating between signal and ASE noise from the optical amplifiers. Both beating contributions build up during the propagation in the physical link (LINK box in Fig. \ref{fig:scheme}) and are then either reduced or enhanced in the virtual link at the receiver-side (DBP box in Fig.~\ref{fig:scheme}). 
\\[12pt]
As described in the following sections, the NLC split ratio that minimizes the signal-noise interaction is different in the case of TRX noise beating and ASE noise beating. ASE noise beating can be typically neglected for short distances which we refer to as the transceiver noise beating regime. On the other hand, TRX noise beating can be neglected for very large distances which we refer to as the ASE noise beating regime. In the following, both regimes are studied separately with respect to their split NLC gains and approximate inequalities are derived that define both regimes.

\subsection{The impact of transceiver noise beating}
\label{sec:TRX_beat_regime}
We define the transceiver noise beating regime as the regime, where the TRX noise beating is much stronger than the ASE noise beating at optimal launch power $\left(\kappa\xi_{\textnormal{TRX}}P_{\text{opt}} \gg \xi_{\textnormal{ASE}}P_{\textnormal{ASE}}\right)$. In the TRX noise beating regime the general SNR \eqref{eq:SNR_general} reduces to
\begin{align}
\text{SNR}=\frac{P}{\kappa{}P+NP_{\textnormal{ASE}}+3\eta\kappa\xi_{\textnormal{TRX}}P^3}.
\label{eq:SNR_TRX_beating_regime}
\end{align}
The optimal NLC split $X_{\text{opt}}$ is obtained by setting the derivative of \eqref{eq:SNR_TRX_beating_regime} with respect to $X$ to zero and solving for $X_{\text{opt}}$. The optimal split is found as
\begin{align}
X_{\text{opt}}=\ninf{\frac{N}{1+\left(\frac{1-\kappa_{\text{R}}}{ \kappa_{\text{R}}}\right)^{\frac{1}{\epsilon}}}},
\label{eq:x_opt_trx}
\end{align}
with the optimal TRX noise beating accumulation factor
\begin{align}
\begin{split}
\xi_{\textnormal{TRX,opt}}&=\frac{\left(1-\kappa_{\text{R}}\right)^{-\frac{1}{\epsilon}}+\kappa_{\text{R}}^{-\frac{1}{\epsilon}}}{\left[\left(1-\kappa_{\text{R}}\right)^{-\frac{1}{\epsilon}}+\kappa_{\text{R}}^{-\frac{1}{\epsilon}}\right]^{1+\epsilon}} \cdot N^{1+\epsilon},
\label{eq:xi_opt_trx}
\end{split}
\end{align}
where $\ninf{x}$ denotes the nearest integer function. This function is the result of the quantization of the number of spans. In the following this rounding is removed for notational convenience. It should be noted that the optimal NLC split ratio $\frac{X_{\text{opt}}}{N}$ is only a function of the transceiver noise ratio and the coherence factor. 
\par 
For comparison, the gain in reach with respect to DBP ($X=0$) is analyzed. The TRX noise accumulation factor for DBP is $\xi_{\textnormal{TRX,DBP}}=\kappa_{\text{R}}N^{1+\epsilon}$ \cite{Galdino_2017_otl}. Inserting $\xi_{\textnormal{TRX,opt}}$ and $\xi_{\textnormal{TRX,DBP}}$ in \eqref{eq:SNR_TRX_beating_regime}, forcing $\text{SNR}_{\textnormal{opt}}=\text{SNR}_{\textnormal{DBP}}$ and solving for $\Delta N_\text{max}=\frac{N_\text{opt}}{N_\text{DBP}}$ yields the reach increase of split NLC with respect to DBP. The result is
\begin{align}
\begin{split}
\Delta N_\text{max} =\left\{ \frac{\kappa_{\text{R}}\left[\left(1-\kappa_{\text{R}}\right)^{-\frac{1}{\epsilon}}+\kappa_{\text{R}}^{-\frac{1}{\epsilon}}\right]^{1+\epsilon}}{\left(1-\kappa_{\text{R}}\right)^{-\frac{1}{\epsilon}}+\kappa_{\text{R}}^{-\frac{1}{\epsilon}}}\right\}^{\frac{1}{3+\epsilon}}.
\label{eq:delta_N}
\end{split}
\end{align}
Similar to the optimal NLC split ratio, the gain in reach is only dependent on the transceiver noise ratio and the coherence factor. Eq.~\eqref{eq:delta_N} yields the gain with respect to DBP. In order to obtain the reach gain compared to DPC ($X=N$), $\kappa_{\text{R}}$ must be replaced by $1-\kappa_{\text{R}}$.
\par
Typical transmission systems in optical communications exhibit a high dispersion coefficient and wide optical bandwidths that result in a small coherence factor. For dispersion parameters $D>16$~$\frac{\text{ps}}{\text{km}\cdot\text{nm}}$, attenuation coefficients $\alpha >0.2$~$\frac{\text{dB}}{\text{km}}$, and optical bandwidths $>100$~GHz, the coherence factor is $\epsilon <0.1$ for $80$~km spans and EDFA amplification \cite[Fig. 10]{Poggiolini_2012_tgm}. Coherence factors for backward pumped Raman-amplified systems are slightly higher yielding $\epsilon <0.17$ for the same parameters \cite[Fig. 3]{Semrau_2017_ace}. For $\epsilon \ll 1$ the optimal NLC split reduces to
\begin{align}
X_{\text{opt}}=\begin{cases}
    0 & \text{if } \kappa_{\text{R}} < 0.5,\\
    \frac{N}{2} & \text{if } \kappa_{\text{R}} = 0.5,\\
    N & \text{if } \kappa_{\text{R}} > 0.5,
  \end{cases}
  \label{hi}
\end{align}
and the TRX noise beating accumulation factor reduces to
\begin{align}
\xi_{\textnormal{TRX,opt}}=\begin{cases}
    \frac{1}{2^{1+\epsilon}}\cdot N^{1+\epsilon} & \text{if } \kappa_{\text{R}} = 0.5,\\
    \text{min}\left[1-\kappa_{\text{R}},\kappa_{\text{R}}\right]\cdot N^{1+\epsilon} & \text{otherwise.}
  \end{cases}
\end{align}
Eq. \eqref{hi} shows that transmission systems with low coherence factors and higher transmitter noise than receiver noise should deploy transmitter-side NLC for maximum performance and vice versa when there is more transmitter noise. In other words, the virtual link should be placed at the one end where less noise is injected. This, perhaps  surprising, result is due to the fact that only transceiver noise beating is considered in this section.
\par 
The split NLC gain in reach with respect to DBP for $\epsilon \ll 1$ yields 
\begin{align}
\Delta N_\text{max}=\begin{cases}
    1 & \text{if } \kappa_{\text{R}} \leq 0.5,\\
    \left(\frac{\kappa_{\text{R}}}{1-\kappa_{\text{R}}}\right)^{\frac{1}{3+\epsilon}} & \text{if } \kappa_{\text{R}} > 0.5 .
  \end{cases}
  \label{eq:delta_N_apx}
\end{align}
There is no split NLC reach gain with respect to DBP for $\kappa_{\text{R}}<0.5$ as DBP is already the optimum itself. When $\kappa_{\text{R}}$ is replaced by $1-\kappa_{\text{R}}$, \eqref{eq:delta_N_apx} gives the split NLC reach gain with respect to DPC due to symmetry reasons. It is apparent from \eqref{eq:delta_N_apx} that transmission systems with low coherence and equally divided transceiver noise ($\kappa_{\text{R}}=0.5$) exhibit no gain compared to DBP in the TRX noise beating regime. However, split NLC gains are significant, when the transceiver noise is unequally divided between transmitter and receiver (e.g. in the case that the ADC introduces more noise than the DAC).  
\begin{figure}
\includegraphics[]{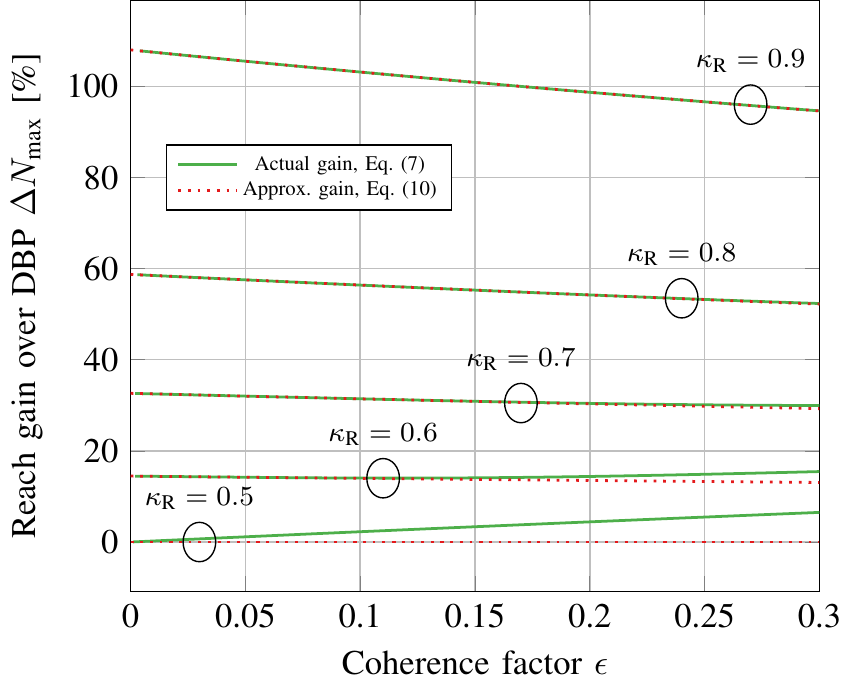}
\caption{Gain in reach of split NLC with respect to DBP as function of the coherence factor for a variety of transceiver noise ratios. Shown are the exact gain from Eq. (7) and its approximation for small $\epsilon$ from Eq. (10).}
\label{fig:reachgain}
\end{figure}
\par
The split NLC reach gain with respect to DBP (Eq. \eqref{eq:delta_N} and its approximation Eq. \eqref{eq:delta_N_apx}) are shown in Fig. \ref{fig:reachgain} as a function of coherence factor for a variety of transceiver noise ratios. Only transceiver ratios $\kappa_{\text{R}}\geq 0.5$ are shown. For lower transceiver noise ratios the plot can be interpreted as the split NLC gain with respect to DPC when $\kappa_{\text{R}}$ is replaced by $1-\kappa_{\text{R}}$. Fig. \ref{fig:reachgain} is sufficient to estimate whether the coherence factor can be considered small and the approximation \eqref{eq:delta_N_apx} can be used. Eq. \eqref{eq:delta_N_apx} serves as an excellent approximation for most of the cases except for high coherence factors combined with a transceiver noise ratio close to $0.5$. The plot also shows that the split NLC reach gain is larger for systems with a larger unbalance between the amount of noise injected by transmitter and receiver. For example, when more noise is injected at the receiver and $\epsilon \ll 1$, the receiver noise beating (occurring in the DBP box in Fig. 1) can be fully removed by placing the complete virtual link at the transmitter. This will result in transmitter noise beating (occurring in the physical link) which will be smaller than the removed receiver noise beating.
\par
In the following, a simple inequality is derived to determine whether a transmission system is operated in the TRX noise beating regime. First, we start with the condition that ASE noise beating is negligible compared to TRX noise beating at optimal launch power
\begin{align}
\xi_{\textnormal{ASE}}P_{\textnormal{ASE}}\ll{}\xi_{\textnormal{TR}}\kappa P_{\textnormal{opt}}.
\label{eq:asebeatin_less_rx}
\end{align}
Inequality \eqref{eq:asebeatin_less_rx} is then expanded as
\begin{align}
\xi_{\textnormal{ASE}}P_{\textnormal{ASE}}\leq \xi_{\textnormal{ASE,DBP}}P_{\textnormal{ASE}}\ll\xi_{\textnormal{TR,opt}}\kappa P_{\textnormal{opt}}\leq{}\xi_{\textnormal{TR}}\kappa P_{\textnormal{opt}},
\label{eq:eq14}
\end{align}
with $\xi_{\textnormal{ASE,DBP}}=\sum_{i=1}^{N}i^{1+\epsilon}$. It is sufficient to consider the inner inequality in \eqref{eq:eq14} in order to show that \eqref{eq:asebeatin_less_rx} holds, which yields (cf. \cite[Appendix]{Galdino_2017_otl})
\begin{align}
\text{SNR}_{\textnormal{EDC,ideal}}\left[\text{dB}\right]
\gg \frac{2}{3}\left(\frac{\text{SNR}_{\textnormal{TRX}}}{\text{min}\left[1-\kappa_{\text{R}},\kappa_{\text{R}}\right]}\right)\left[\text{dB}\right]-6.5\text{dB},
\label{eq:signa-TRX_regime}
\end{align}
where $\left(\cdot\right)\left[\text{dB}\right]$ means conversion to decibel scale and $\text{SNR}_{\textnormal{EDC,ideal}}$ is the SNR at optimal launch power with electronic dispersion compensation only and no transceiver noise, which can be calculated as
\begin{align}
\text{SNR}_{\textnormal{EDC,ideal}}=\frac{1}{\sqrt[3]{\frac{27}{4}P_{\textnormal{ASE}}^2\eta{}N^{3+\epsilon}}}.
\label{eq:maxSNREDC}
\end{align}
When inequality \eqref{eq:signa-TRX_regime} is satisfied, the corresponding system is operating in the transceiver noise beating regime and the optimal split ratio and reach gain reported in this section applies.

\subsection{The impact of ASE noise beating}
\label{ASE noise beating regime}
In this section the regime is discussed where the TRX noise beating is much weaker than ASE noise beating at optimal launch power $\left(\kappa\xi_{\textnormal{TRX}}P_{\text{opt}} \ll \xi_{\textnormal{ASE}}P_{\textnormal{ASE}}\right)$. This regime has already been studied in the literature \cite{Lavery_2016_tbo,Lavery_2017_otb,Ellis_2015_clo} and is therefore only briefly covered. In the ASE noise beating regime the general SNR \eqref{eq:SNR_general} reduces to
\begin{align}
\text{SNR}=\frac{P}{NP_{\textnormal{ASE}}+3\eta\xi P^2P_{\textnormal{ASE}}},
\label{eq.SNREDC}
\end{align}
with the optimal NLC split given as $X_{\textnormal{opt}}=\ceil*{\frac{N}{2}}$, where $\ceil*{x}$ denotes the ceiling function with the optimal ASE noise beating accumulation factor \cite[Eq. (7)]{Lavery_2016_tbo}
\begin{align}
\xi_{\textnormal{ASE,opt}}=\left(\frac{N}{2}\right)^{1+\epsilon}+2\sum_{i=1}^{\frac{N}{2}-1}i^{1+\epsilon}.
\end{align}
Similar to section \ref{sec:TRX_beat_regime}, the gain of split NLC is compared to the performance of DBP. The gain in reach $\Delta N_\text{max}=\frac{N_\text{opt}}{N_\text{DBP}}$ can be expressed as 
\begin{align}
\begin{split}
\Delta N_\text{max} = 2^{\frac{1+\epsilon}{3+\epsilon}}.
\label{eq:delta_N_ASE_apx}
\end{split}
\end{align}
The split NLC reach increase is only a function of the coherence factor with $\Delta N_\text{max} \to 25$\% for $\epsilon \ll 1$. This means that a reach increase of $25$\% is expected for typical high bandwidth transmission systems in optical fiber communications.
\par 
Similarly to section \ref{sec:TRX_beat_regime}, an inequality is derived to determine whether a transmission system is operated in the ASE noise beating regime. First, we start with the condition that TRX noise beating is negligible compared to ASE noise beating  
\begin{align}
\xi_{\textnormal{ASE}}P_{\textnormal{ASE}}\gg{}\xi_{\textnormal{TR}}\kappa P_{\textnormal{opt}},
\label{eq:a}
\end{align}
which is then expanded to (for $\epsilon \ll 1$)
\begin{align}
\begin{split}
& \xi_{\textnormal{ASE}}P_{\textnormal{ASE}}>\xi_{\textnormal{ASE,opt}}P_{\textnormal{ASE}}\approx \frac{1}{2}\xi_{\textnormal{ASE,DBP}}P_{\textnormal{ASE}} \\
& \gg\xi_{\textnormal{TRX,max}}\kappa P_{\textnormal{opt}}>{}\xi_{\textnormal{TRX}}\kappa P_{\textnormal{opt}}.
\end{split}
\end{align}
Considering only the inner inequality to prove \eqref{eq:a} yields
\begin{align}
\text{SNR}_{\textnormal{EDC,ideal}}\left[\text{dB}\right]
\ll \frac{2}{3}\left(\frac{\text{SNR}_{\textnormal{TRX}}}{\text{max}\left[1-\kappa_{\text{R}},\kappa_{\text{R}}\right]}\right)\left[\text{dB}\right]-9.5\text{dB},
\label{eq:ineq17}
\end{align}
with $\text{SNR}_{\textnormal{EDC,ideal}}$ as in \eqref{eq.SNREDC}. When inequality \eqref{eq:ineq17} holds, the corresponding transmission system is operated in the ASE noise beating regime and the optimal split ratio and the reach gain reported in this section applies. 
\section{Simulation Results}
\label{sim}
In this section two optical transmission systems are simulated by numerically solving the Manakov equation using the split-step Fourier (SSF) algorithm with parameters listed in Table~\ref{tab:parameters}. Additive white Gaussian noise was added at transmitter and receiver to emulate a finite transceiver SNR and nonlinearity compensation was carried out as schematically shown in Fig.~\ref{fig:scheme}. A matched filter was used to obtain the output symbols and the SNR was ideally estimated as the ratio between the variance of the transmitted symbols $E[|X|^2]$ and the variance of the noise $\sigma^2$, where $\sigma^2=E[|X-Y|^2]$ and $Y$ represents the received symbols after digital signal processing. The nonlinear interference coefficient and the coherence factor were obtained in closed-form from \cite[Eq. (13) and Eq. (23)]{Poggiolini_2012_tgm} with the modulation format dependent correction from \cite[Eq. (2)]{Poggiolini_2015_asa}. Closed-form expressions for both quantities in the context of Raman amplification can be found in \cite{Semrau_2017_ace}.
\begin{center}
\captionof{table}{Simulation Parameters}
\label{tab:parameters}
  \begin{tabular}{ l | l l l }
    \hline
   \textbf{Parameters} &  \\ \hline
  Span length [km]& 80 \\ \hline  
  Loss ($\alpha$) [dB/km]& 0.2 \\ \hline
    Dispersion ($D$) [ps/nm/km]& 17 \\ \hline
    NL coefficient ($\gamma$) [1/W/km]& 1.2\\ \hline
      Number of channels & 3 \\ \hline
        Optical bandwidth ($B$) [GHz]& 96\\ \hline
    Symbol rate ($R_b$) [GBd]&   32  \\ \hline
    Channel spacing [GHz]& 32 (Nyquist)\\ \hline
    Roll-off factor [\%]& 0 \\ \hline
    Noise Figure [dB]& 4 \\ \hline
    Transceiver SNR ($\text{SNR}_{\textnormal{TRX}}$) [dB]& 26 \\ \hline
       Oversampling & 3 \\ \hline
         Number of SSF steps per span & 800 log-distributed \\ \hline
          Modulation format & 256-QAM \\ \hline
        NLI coeff. ($\eta$) [dB$\left(\text{1/}\text{W}^2\right)$]&   26.2  \\ \hline
        Coherence factor ($\epsilon$) &   0.108  \\ \hline
  \end{tabular}
\end{center}
In order to test the theory presented in section \ref{sec:theory}, a system with a transceiver noise that is equally divided between transmitter and receiver ($\kappa_{\text{R}}=0.5$) and a system with an unequal division of transceiver noise ($\kappa_{\text{R}}=0.8$) are 	simulated.

\subsection{Equal transmitter and receiver noise contribution}
\begin{figure*}
\centering
\includegraphics[]{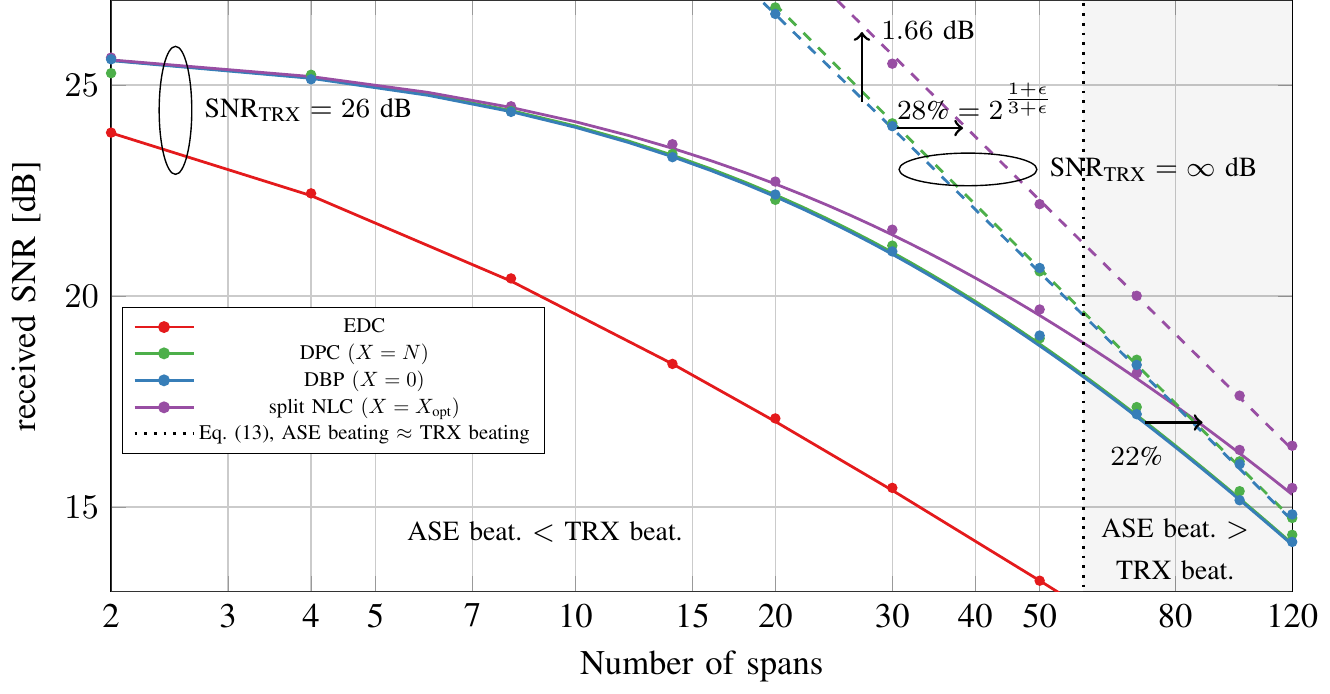}
\caption{SNR at optimum launch power as a function of span number obtained by simulation (markers) and Eq. (1) (lines). The case with an infinite transceiver SNR (solid lines) and a finite transceiver SNR of $26$ dB (dashed lines) are shown. The transceiver noise is equally divided between transmitter and receiver ($\kappa_{\text{R}}=0.5$).}
\label{fig:SNR_distance_kr_05}
\end{figure*}
An optical transmission system with an equal share of transceiver noise between transmitter and receiver ($\kappa_{\text{R}}=0.5$) is simulated. The received SNR as a function of distance is shown in Fig. \ref{fig:SNR_distance_kr_05}. The lines represent the analytical model estimated by \eqref{eq:SNR_general} at optimal launch power for electronic dispersion compensation (EDC), DBP ($X=0$), DPC ($X=N$) and split NLC with the optimal split NLC of $X_{\text{opt}}=\ceil*{\frac{N}{2}}$ between transmitter and receiver. A split of $X=\ceil*{\frac{N}{2}}$ is the optimum for a system where the transceiver noise is equally divided between transmitter and receiver. For the EDC case a summand $\eta^{1+\epsilon}P^3$ is added in the denominator to include signal-signal nonlinearity in \eqref{eq:SNR_general}. Markers represent results obtained by numerical simulation. Furthermore, the same transmission system without transceiver noise ($\text{SNR}_{\text{TRX}}=\infty$ dB) is shown with dashed lines and the point where ASE noise beating approximately equals TRX noise beating is shown with a black vertical dashed line. For the given system parameters, both beating contributions are approximately equal at $58$ spans according to \eqref{eq:signa-TRX_regime}. 
\par 
The model is in excellent agreement with the simulation results. Fig. \ref{fig:SNR_distance_kr_05} shows that transceiver noise significantly reduces the gains of nonlinearity compensation compared to EDC. In the case of finite transceiver SNR, the NLC gains increase with distance, as the transceiver SNR has less impact for lower values of received SNR. This is contrast to the case of no transceiver noise, where the gains of nonlinear compensation decrease with distance. 
\par 
Further, in the case of a finite transceiver SNR, there is negligible performance difference between DPC (green line) and DBP (blue line), as they only differ for short distances due to an advantage of one span in favor of DPC in the ASE noise beating contribution \cite[Fig. 2]{Lavery_2016_tbo}. However, short transmission distances are dominated by TRX noise beating where both perform the same (cf. Eq. \eqref{eq:xitrx} with $\kappa_{\text{R}}=0.5$).
\par
Moreover, as predicted in Section \ref{sec:theory}, there is negligible gain of split NLC when TRX noise beating is dominant. The left-hand side of \eqref{eq:signa-TRX_regime}, which defines the TRX noise beating regime, scales as $-10$ dB per decade in distance increase (for $\epsilon \ll 1$). Both beating contributions are approximately equal at 58 spans for the chosen parameters. Therefore, the TRX noise beating contribution is 10~dB higher than the ASE noise beating at $\frac{58}{10} \approx 6$ spans and the transmission system is well inside the TRX noise beating regime. At this point, ASE beating starts to be notable and split NLC begins to yield notable gains.
\par 
At $72$ spans the gain of split NLC in reach compared to DBP is $22$\%. Even at such a long transmission distance, the gain is not fully converged to the case of $\text{SNR}_{\text{TRX}}=\infty$~dB. According to \eqref{eq:ineq17}, a span number of at least $580$ is required for the TRX noise beating to be one order of magnitude lower than ASE noise beating. Such distances are not of practical interest, which illustrates the importance of transceiver noise beating in real systems. Inequality \eqref{eq:ineq17} can further be used to estimate the impact of a different transceiver SNR. As an example, to shift the point where ASE noise beating approximately equals TRX noise beating to $5.8$ spans, a transceiver SNR of an extraordinary $41$~dB would be needed. Both calculations underline the importance of TRX noise beating in relation to ASE beating.
\par 
Fig. \ref{fig:SNR_distance_kr_05} shows that realistic systems are usually operated in the TRX noise beating regime for short, medium and long-haul distances and in a mixed regime for transatlantic and transpacific distances. Split NLC proves only useful in the latter case for transmission systems with \textit{equally} divided transceiver noise. 
\subsection{Unequal transmitter and receiver noise contribution}
\begin{figure*}
\centering
\includegraphics[]{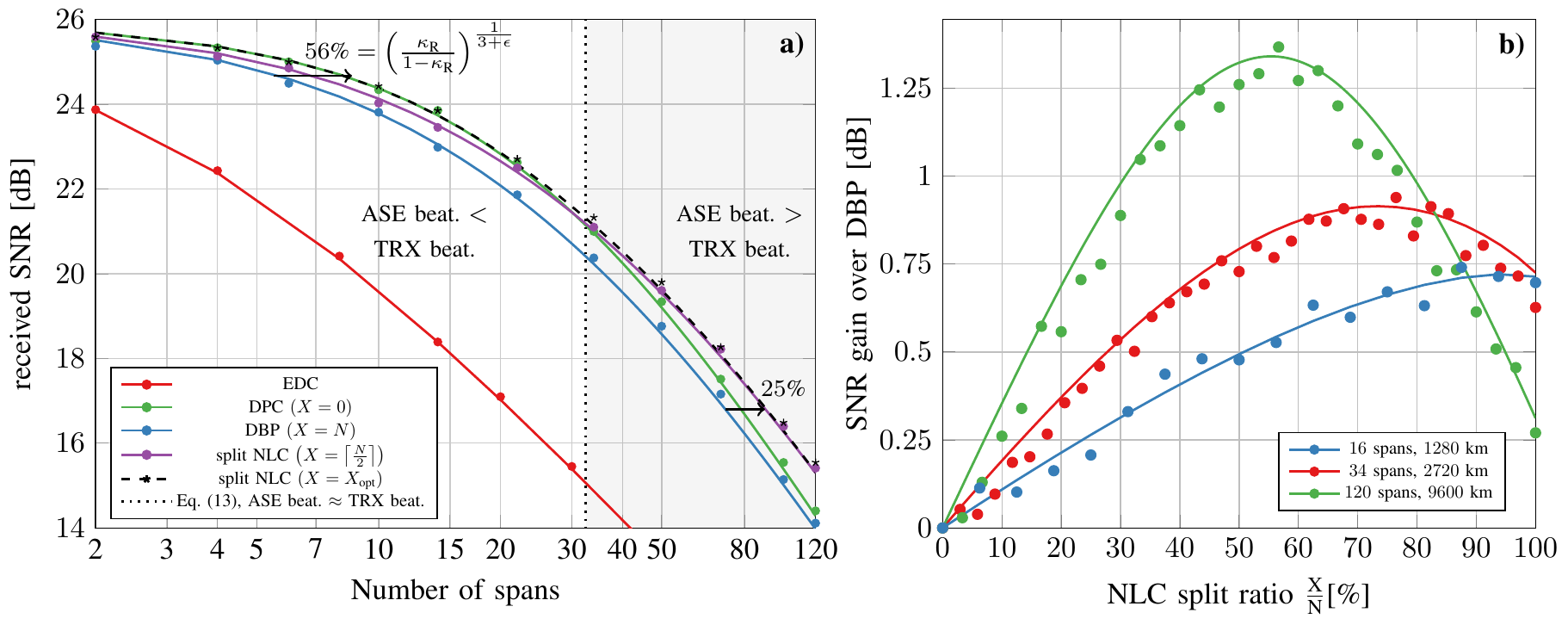}
\caption{a) SNR at optimum launch power as a function of span number and the SNR as a function of NLC split ratio b) obtained by simulation (markers) and Eq. (1) (lines). The transceiver SNR is $26$~dB and the transceiver noise is unequally divided between transmitter and receiver ($\kappa_{\text{R}}=0.8$).}
\label{fig:SNR_distance_kr_08}
\end{figure*}

In this section the same optical transmission system as in the previous section is simulated but with 20\% of the transceiver noise injected at the transmitter and 80\% injected at the receiver ($\kappa_{\text{R}}=0.8$). Unequal contributions of transceiver noise are more likely in realistic transmission systems. The SNR at optimal launch power as a function of distance is shown in Fig. \ref{fig:SNR_distance_kr_08}a). The lines represent the analytical model estimated by \eqref{eq:SNR_general} at optimal launch power for EDC, DPC and DBP. Further, a NLC split of $X=\ceil*{\frac{N}{2}}$ and the optimal split $X_{\text{opt}}$ obtained by taking the maximum of all possible splits $X \in \left[0,N\right]$ are shown. 
The absolute SNR as well as the SNR gain predictions of the model are in very good agreement with the simulation results. Fig. \ref{fig:SNR_distance_kr_08}a) shows that optimal split NLC yields significant reach gain with respect to DBP throughout all distances. For instance, in the TRX noise beating regime a reach gain of 56\% is achieved (from 5 to 8 spans). This is in stark contrast to the case of equal division of transceiver noise in the previous section and confirms the theory presented in section \ref{sec:theory}. In the TRX noise beating regime the optimal NLC split is $X=N$ which is equivalent to the DPC case. As shown in Fig. \ref{fig:SNR_distance_kr_08}a) DPC performs optimally up to approximately 30 spans where the amount of ASE noise beating becomes comparable to the amount of TRX noise beating. As the coherence factor is quite low ($\epsilon = 0.108$), the simple Eq.~\eqref{eq:delta_N_apx} accurately predicts the reach gain, yielding a reach increase of $56$\% for this example. The DPC curve starts to approach the DBP curve with a residual gap as the TRX noise beating is not negligible up to this point. Consequently, the optimal NLC split ratio at $120$ spans is 56\% with a gain of $1.34$~dB in SNR with respect to DBP. Those gains are still not in-line with the theoretical results in section \ref{ASE noise beating regime} as some residual transceiver noise still affects the transmission.
\par 
The split NLC gain with respect to DBP as a function of the NLC split ratio is shown in Fig. \ref{fig:SNR_distance_kr_08}b) for $16$, $34$ and $120$ spans. The gain of optimal split NLC is $0.74$~dB at 16 spans. As the ASE noise beating becomes more significant, the optimal split ratio slowly shifts from $X=N$ to $X=\ceil*{\frac{N}{2}}$. At $34$ spans, where the amount of ASE noise beating is approximately equal to the amount of TRX noise beating, the relative optimal NLC split ratio is $73$\%. For longer distances the ASE noise beating contribution becomes more dominant and the optimal NLC split ratio is with $54$\% close to $X=\ceil*{\frac{N}{2}}$. 
\par 
It might be surprising for the reader that the gain in reach is decreasing with transmission distance (e.g., from $56$\% at 5 spans to $25$\% at 70 spans) but the gain in SNR is increasing with distance (e.g., from $0.4$~dB at 5 spans to $1.1$~dB at 70 spans). Split NLC seems to yield higher SNR gains for longer distances and higher reach gains for shorter distances. This effect is due to the linear transceiver noise term $\kappa P$ in Eq.~\eqref{eq:SNR_general}. Different received SNRs are affected differently by the linear transceiver noise contribution and as a result the SNR gains for short distances are not visible. Hence, the gain in SNR as a figure of merit may be a misleading quantity to compare nonlinearity compensation techniques in systems that are impaired by transceiver noise. A better figure of merit is reach increase evaluated at the same received SNR, as the linear transceiver noise affects both points equally. 
\begin{figure}
\centering
\includegraphics[]{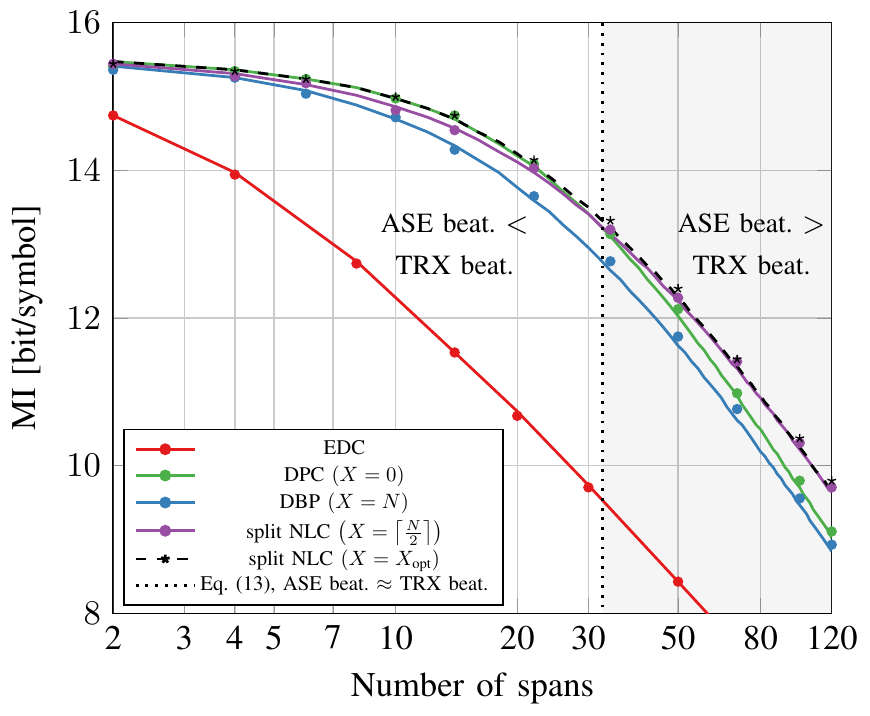}
\caption{MI at optimum launch power as a function of span number obtained by simulation (markers) and Eq. (1) (lines). A finite transceiver SNR of $26$~dB is assumed where the transceiver noise is unequally divided between transmitter and receiver ($\kappa_{\text{R}}=0.8$).}
\label{fig:MI_distance_kr_08}
\end{figure}
\par 

The obtained SNR values were further used to estimate the mutual information (MI) per polarization as described in \cite{Maher_2015_reo}. Mutual information is defined as
\begin{equation}
\text{MI} = \frac{1}{m}\sum_{x \in \mathbb{X} }\int_{ \mathbb{C}}p(y|x)\log_{2}\frac{p(y|x)}{p(y)}dy,
\label{Eq:MI}
\end{equation}
where $m$ is the cardinality of the QAM constellation, $x$ and $y$ are random variables representing the transmitted and received symbols and $\mathbb{X}$ is the set of possible transmitted symbols. Assuming an additive white Gaussian noise (AWGN) channel, the channel law is given by
\begin{equation}
p(y|x)=\frac{1}{\pi\sigma^2_{n}}\exp\left(-\frac{|y-x|^2}{\sigma^2_{n}}\right).
\label{Eq:AWGN_Channel}
\end{equation}     
Eq.~\eqref{Eq:MI} was then numerically integrated using the Monte-Carlo method. The resulting MI of the simulated transmission system is shown in Fig.~\ref{fig:MI_distance_kr_08} as a function of span number. At $34$ spans (1920~km) DBP increases the MI compared to EDC by $3.2$ bit/symbol and split NLC yields another $0.55$ bit/symbol on top of the DBP gain. This shows that a reasonable throughput increase can be achieved by applying split NLC.
\section{Conclusion}
The performance of split nonlinearity compensation was analyzed in the context of realistic transceiver sub-systems. It was demonstrated that the gain of split NLC and the optimal split ratio are strongly dependent on whether TRX noise or ASE noise beating dominates. Simple formulas were derived that can be used for system design and gain prediction. It was found that split NLC yields negligible gain compared to DBP for distances below 800~km, when the transceiver noise is equally distributed between transmitter and receiver. However, when the transceiver noise is unequally distributed, reach increases of 56\% on top of digital back-propagation are achievable for the system under test. Alternatively, split NLC can be applied to increase the mutual information by 0.55 bits/symbol for distances larger than 1440~km, compared to DBP. This demonstrates that significant throughput or reach increase can be achieved by properly tailoring the digital nonlinearity compensation algorithm to the noise distribution of the underlying optical transmission system. The results of this work suggest that split NLC yields greater reach increases than current experimental demonstrations using DBP or DPC. This demonstration is left for future work.

\section*{Acknowledgment}

Financial support from UK EPSRC programme grant UNLOC (EP/J017582/1) and a Doctoral Training Partnership (DTP) studentship to Daniel Semrau is gratefully acknowledged. The authors thank G. Liga from University College London for valuable comments on previous drafts of this paper.

\ifCLASSOPTIONcaptionsoff
  \newpage
\fi

\bibliographystyle{IEEEtran}
\bibliography{IEEEabrv,ref}

\end{document}